\documentclass[twocolumn,showpacs,preprintnumbers,amsmath,amssymb]{revtex4}

\usepackage{graphicx}
\usepackage{dcolumn}
\usepackage{bm}

\begin{document}

\preprint{APS/123-QED}

\title{Electron-Transport Properties of Na Nanowires under Applied Bias Voltages}

\author{Shigeru Tsukamoto}%
\affiliation{%
Department of Precision Science and Technology, Osaka University, Suita, Osaka 565-0871, Japan
}%

\author{Kikuji Hirose}%
\affiliation{%
Department of Precision Science and Technology, Osaka University, Suita, Osaka 565-0871, Japan
}%

\date{\today}

\begin{abstract}
We present first-principles calculations on electron transport through Na nanowires at finite bias voltages. The nanowire exhibits a nonlinear current-voltage characteristic and negative differential conductance. The latter is explained by the drastic suppression of the transmission peaks which is attributed to the electron transportability of the negatively biased plinth attached to the end of the nanowire. In addition, the finding that a voltage drop preferentially occurs on the negatively biased side of the nanowire is discussed in relation to the electronic structure and conduction.
\end{abstract}

\maketitle

Currently, quantized conductances of atomic nanowires with a few atoms width and nanometer length, which are in steps of $G_0=2e^2/h$ ($e$: the electron charge, $h$: Planck's constant), have attracted much attention because of both fundamental and technological importance. A number of exciting experimental and theoretical studies on the conductance quantization for various metal nanowires have been reported\cite{JMRuitenbeek00,JMKrans95,HOhnishi98,STsukamoto01,NKobayashi00,ANakamura99,NDLang97,NDLang95,RNBarnett97,JCCuevas98,HSSim01}. Above all, the studies for Na nanowires are of particular interest, because the nature of electron conduction through nanowires is expected to emerge best for Na nanowires having the simplest electronic structure where monovalent electrons predominate. Elaborate investigations on electron-transport properties of Na nanowires are, however, not many yet. A few experiments have been reported\cite{JMKrans95}, where the variation of the conductance was measured during the withdrawal of the Na film to show the stairlike behavior of the quantized conductance. Theoretically, there have been some first-principles calculations fully self-consistently performed using the model incorporating a couple of semi-infinite electrodes so as to evaluate electron flow through Na nanowires\cite{STsukamoto01,NKobayashi00,ANakamura99,NDLang97}. But, these theoretical studies are limited to zero bias voltage\cite{ANakamura99,NKobayashi00,NDLang97} or the low bias regime\cite{STsukamoto01}. The analysis and discussion of electron transport through the Na nanowire should be extended to a higher bias voltage in order to gain insight into a unique conduction property possible under a strong electric field.

In this letter, we present a theoretical analysis of electron transport through the Na nanowire under the application of a sizable bias voltage, based on the density functional theory within the local density approximation. All the calculations determining the electronic structure and transport properties are self-consistently evaluated by the method making use of the Lippmann-Schwinger equation proposed by Lang\cite{NDLang95}. The norm-conserving pseudopotentials\cite{NTroullier91} are adopted to describe the electron-ion interaction, and the real-space timesaving double-grid technique\cite{TOno99} is used by virtue of an efficient computation of the nonlocal parts of the pseudopotentials. We will demonstrate that the Na nanowire exhibits an unusual nonlinear current-voltage (I-V) characteristic which leads to negative differential conductance behavior. Notice that the negative differential conductance is expected to be an essential property to realize fast switching in future electronic devices. This paper is probably the first report describing that a simple nanowire in which all atoms are tightly binding together possesses the negative feature of the differential conductance\cite{comment2}. We will also discuss how an applied bias voltage gives rise to the change of the effective potential along the nanowire (voltage drop) and of the charge density distribution.

According to Refs.~\onlinecite{NKobayashi00} and \onlinecite{STsukamoto01}, we here adopt the model composed of three parts for the nanowire system: a straight atomic nanowire, two plinths, and a couple of semi-infinite jellium electrodes sandwiching them (see Fig.~\ref{fig:Figure6} for illustration). The straight nanowire is made up of $N$ ($N$=1$\sim$5) Na atoms equispaced at 7.0 a.u., which is equal to the nearest neighbor atomic distance of the Na crystal. Each plinth is modeled after the Na(001) ideal surface, i.e., it is formed by four Na atoms, one at each corner of a square with the side length of 8.1 a.u. These plinths are attached to the two ends of the straight nanowire and act as buffer layers to moderate the unphysical effect due to an oversimplified electrode model of jellium. The electron density of the jellium electrodes is taken equal to the average valence-electron density of the Na crystal ($r_s=4.0$ a.u.)\cite{NKobayashi00}. The distance between the jellium edge and the atoms forming the plinth is set to be 4.0 a.u. In the present calculation, we do not consider any structural relaxation.

In the numerical calculations, we employ the supercell on which periodic boundary conditions are imposed in the directions perpendicular to the nanowire ($x$ and $y$ directions) and nonperiodic boundary condition in the direction parallel to it ($z$ direction). Therefore, wave functions are expressed by the Laue representation. The side length of the supercell in the $x$ and $y$ directions is 24.0 a.u. and the number of the plane waves in these directions is taken to be $24^2$, which corresponds to a cutoff energy of 7.4 Ry. We use only the $\Gamma$ point in the $xy$ Brillouin zone because of a large supercell adopted here. The grid size in the $z$ direction is chosen to be 0.5 a.u.

\begin{figure}[tb]
\includegraphics{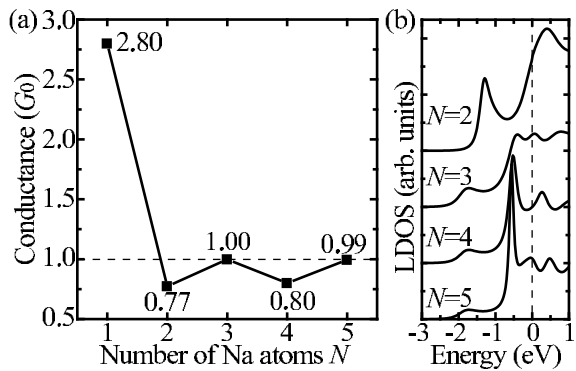}
\caption{\label{fig:Figure1}(a) Conductance and (b) LDOS around the straight nanowire of the respective systems at the zero bias limit. In (b), the broken line represents the Fermi level. The LDOS curves are displaced vertically for clarity.}
\end{figure}
Before moving on to the discussion of electron-transport properties for the Na nanowire system at a finite bias voltage, we briefly report the results obtained at the zero bias limit. In Fig.~\ref{fig:Figure1}(a), the calculated conductance is drawn as a function of the number of atoms $N$ in the straight nanowire: in the case of $N=1$, i.e., the case of the closed packed structure of Na atoms, the conductance takes the value $2.8G_0$, while in the cases of $N=$2$\sim$5, it is close to the quantized value of $1G_0$\cite{NKobayashi00}. These features of the conductance are in good agreement with the experimental results obtained by Krans {\it et~al.}, where the conductance histogram shows the three distinct points: a peak at slightly lower than the $3G_0$ quantized value, a peak at $1G_0$ and a missing $2G_0$ peak\cite{JMKrans95}. In Fig.~\ref{fig:Figure1}(a), an oscillatory behavior in the conductance is observed for $N=2\sim5$, i.e., the even-atom-number nanowire systems exhibit slightly lower conductance than the quantized value of $1G_0$, and the odd-atom-number systems show fully quantized conductance $1G_0$. This oscillation can be understood in terms of the local density of states (LDOS) around the nanowire at the Fermi level: as seen in Fig.~\ref{fig:Figure1}(b), the LDOS for the even-atom-number (odd-atom-number) nanowire has an off-resonance (on-resonance) transport property at the Fermi level\cite{NDLang98}. Based on the Friedel sum rule, Sim {\it et~al.} recently predicted the occurrence of the even-odd oscillation for the Na nanowire sandwiched by the pyramidal-shape tips \cite{HSSim01}. Our present results verify their prediction of the even-odd oscillation.

\begin{figure}[tb]
\includegraphics{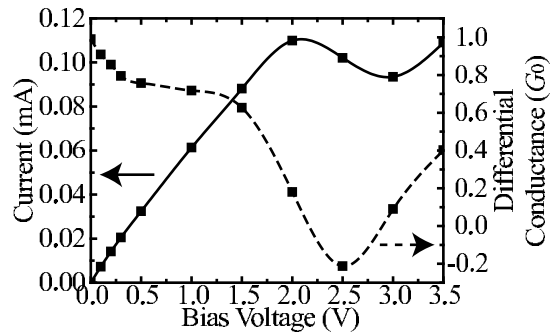}
\caption{\label{fig:Figure2}I-V characteristic of the $N=5$ Na nanowire system (solid line) and its differential conductance (broken line).}
\end{figure}
We now discuss how electron-transport properties of the nanowire are affected when a sizable bias voltage is applied. The range of bias voltages is 0.0$\sim$3.5V and a long nanowire of $N=5$ is adopted. The current per supercell is simply expressed as the integration of the transmission over the voltage window, which represents the energy range between the Fermi levels of the two electrodes. Hereafter, the respective electrodes are called left and right electrodes, and the bias voltage is applied with the left electrode negative and the Fermi level of the right electrode kept fixed. Figure~\ref{fig:Figure2} shows the I-V characteristic of the system and its differential conductance. Here, the differential conductance is defined as the derivative of the current with respect to the bias voltage. It is clearly seen that the system exhibits the nonlinear I-V characteristic and in particular, the differential conductance shows unusual behavior near zero bias voltage and around 2.5 V; the differential conductance rapidly decreases in the range of 0.0$\sim$0.5 V, deviating from the quantized value $1G_0$ observed at the zero bias limit, and moreover, it is negative at the bias voltages of 2.3$\sim$2.7 V with a minimum at 2.5 V.

\begin{figure}[tb]
\includegraphics{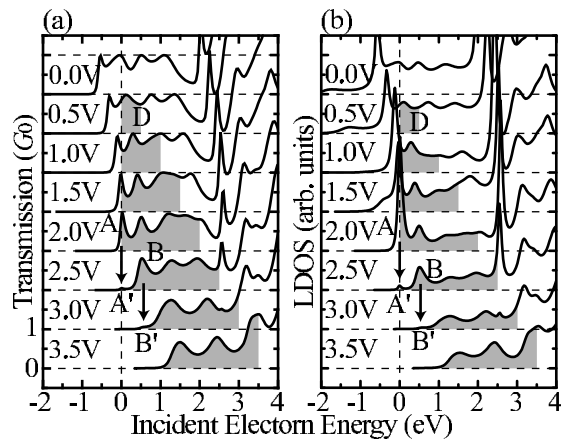}
\caption{\label{fig:Figure3}(a) Transmission and (b) LDOS around the straight nanowire of electrons incident from the left electrode. The energy is measured from the Fermi level of the right electrode. The shaded areas show the voltage windows. The eight curves in each panel correspond to the bias voltages indicated, and are displaced vertically for clarity.}
\end{figure}
We first examine why the reduction of the differential conductance is induced at the low bias voltages of 0.0$\sim$0.5 V. The transmission and the LDOS around the straight nanowire of electrons incident from the left electrode are depicted as a function of the electron energy measured from the Fermi level of the right electrode in Figs.~\ref{fig:Figure3}(a) and (b), respectively. The right edges of the voltage windows (shaded areas) in these figures correspond to the Fermi level of the left electrode for the respective bias voltages. When the bias voltage is boosted from 0.0 V to 0.5 V, a dip in the transmission curve, which is indicated by D in Fig.~\ref{fig:Figure3}(a), appears in the voltage window, causing the reduction in the differential conductance. This transmission dip D is brought about by the off-peak of the LDOS marked D in Fig.~\ref{fig:Figure3}(b)

\begin{figure}[tb]
\includegraphics{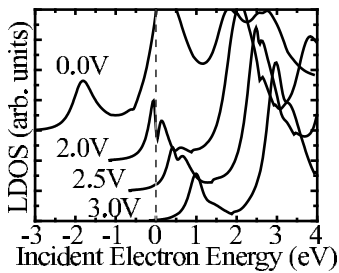}
\caption{\label{fig:Figure4}LDOS around the left plinth at the bias voltages of 0.0, 2.0, 2.5 and 3.0 V. The energy is measured from the Fermi level of the right electrode. The curves are displaced vertically for clarity.}
\end{figure}
Next, we discuss the negative feature of the differential conductance at around 2.5 V seen in Fig.~\ref{fig:Figure2}. In the transmission in Fig.~\ref{fig:Figure3}(a), one can see that the leftmost peak clearly observed at the bias voltage of 2.0 V is suddenly suppressed at 2.5 V (A$\rightarrow$A'), and that one more peak at 2.5 V is also suppressed when the bias voltage is raised to 3.0 V (B$\rightarrow$B')\cite{comment1}. As a consequence of the suppression, the integration of the transmission over the voltage window decreases upon raising the bias voltage from 2.0 V to 3.0 V, which leads to the negative differential conductance. In contrast, such a disappearance of the transmission peak is no longer observed at the bias voltage of 3.5 V. It should be noticed that the suppression of the transmission peaks is in accordance with the vanishing of the corresponding peaks in the LDOS around the straight nanowire of electrons incident from the left electrode, marked by A$\rightarrow$A' and B$\rightarrow$B' in Fig.~\ref{fig:Figure3}(b). This vanishing implies that some electrons coming from the left electrode with specific energies cannot reach the nanowire when the bias voltage exceeds a certain threshold.

\begin{figure}[tb]
\includegraphics{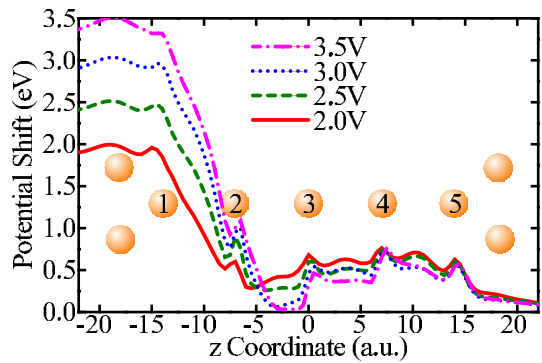}
\caption{\label{fig:Figure5}(color) Shifts of the local effective potential along the axis of the nanowire upon applying the bias voltages of 2.0, 2.5, 3.0 and 3.5 V. The atomic geometry is shown in the panel. The Na atoms forming the $N=5$ straight nanowire are labeled 1$\sim$5.}
\end{figure}

\begin{figure}[tb]
\includegraphics{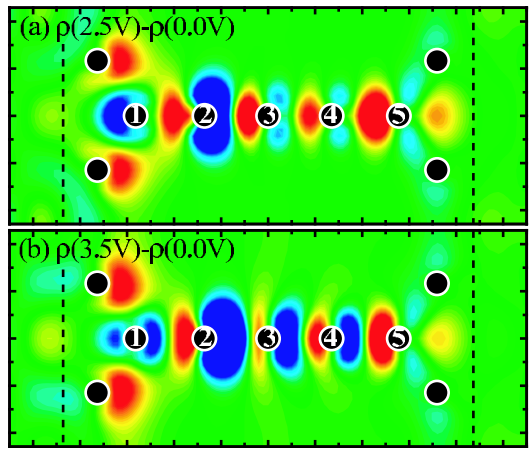}
\caption{\label{fig:Figure6}(color) Changes in the charge density distribution upon applying the bias voltages of (a) 2.5 V and (b) 3.5 V. The red shows the increase and the blue the decrease. An area with sides of 55.0 a.u. $\times$ 23.0 a.u. is displayed. The Na atoms forming the $N=5$ straight nanowire are labeled 1$\sim$5. Broken lines represent the edges of the jellium electrodes.}
\end{figure}

In order to investigate why these transmission peaks and LDOS peaks around the straight nanowire suddenly disappear as the bias voltage is raised, we focus on the electronic structure around the {\it left plinth}. Figure~\ref{fig:Figure4} shows the LDOS around the left plinth of electrons incident from the left electrode at the bias voltages of 0.0, 2.0, 2.5 and 3.0 V, where each curve shifts to the higher energy side according to the applied bias voltage. This shift corresponds to the linear rising of the effective potential around the left plinth upon applying bias voltage, which will be discussed later for Fig.~\ref{fig:Figure5}. We see in Fig.~\ref{fig:Figure4} that the leftmost LDOS peaks appear at 0.0, 0.5 and 1.0 eV at the bias voltages of 2.0, 2.5 and 3.0 V, respectively. It is now common knowledge that the energy of the leftmost LDOS peak corresponds to the {\it first onset energy} of the transmission\cite{NKobayashi00,STsukamoto01,JCCuevas98}. Consequently, when the bias voltage of 2.0 V is applied, electrons with energy lower than 0.0 eV are refused to go through the left plinth, while those with energy above 0.0 eV readily enter the plinth and the nanowire to yield the transmission peak A in Fig.~\ref{fig:Figure3}(a). In the case of the bias voltage of 2.5 V, the threshold energy for electrons to pass through the left plinth is 0.5 eV, therefore, electrons with the energy of 0.0 eV are prohibited from entering the left plinth. Thus, as the bias voltage is boosted from 2.0 V to 2.5 V, the suppression of the transmission peak at 0.0 eV, A$\rightarrow$A', is brought about. This mechanism of the suppression of the transmission peak also holds in the case of B$\rightarrow$B' in Fig.~\ref{fig:Figure3}(a).

Finally, we explore how the applied bias voltages drop along the nanowire. Figure~\ref{fig:Figure5} shows the differences of the local effective potential along the axis of the nanowire at some significant bias voltages from that at zero bias voltage, $V_{eff}(V_B)-V_{eff}(0.0)$, at $V_B=$2.0, 2.5, 3.0 and 3.5 V. In Fig.~\ref{fig:Figure5}, one can see that the intensive voltage drop takes place mainly around the negatively biased end of the nanowire (atoms numbered 1 and 2 in Fig.~\ref{fig:Figure5})\cite{MBrandbyge99}, while the voltage drop is nominal near the two plinths and on the positively biased side (atoms 3$\sim$5). The localization of the voltage drop is understood as follows. Electrons abundant in proportion to the applied bias voltage rush into the left end of the nanowire from the left electrode through the plinth. The congestion of electron flow occurs around the region between atoms 1 and 2, where the narrowing of the path along which electrons pass takes place due to the geometrical transition from the wider electrode to the narrower nanowire. Although only electrons with some specific momenta can enter the conduction channel and reach the opposite electrode without any backscattering, the remaining electrons suffer from a considerable amount of reflection. Thus, the resistance for electron flow localizes around the region between atoms 1 and 2, which results in the intensive voltage drop there. This mechanism of electron flow is analogous to the {\it vena contracta} effect known in hydrodynamics. The absence of a sizable voltage drop near the two plinths and on the positively biased side of the nanowire is associated with the change in the charge density distribution upon applying bias voltage. Figure~\ref{fig:Figure6} shows the profiles of the difference of the charge densities, $\rho(V_B)-\rho(0.0)$, at $V_B=$2.5 and 3.5 V. On the right of the atoms forming the left plinth and the left of atom 2, one observes considerable increases of the charge density. These increases reflect the screening effect which keeps the electric potential of the negatively biased jellium electrode and plinth equal to the negative electrode potential applied to the left jellium. On the other hand, the most significant decreases of electrons occur on the right of atom 2 so that the electric potential of the positively biased side of the nanowire system (atoms 3$\sim$5, the right plinth and jellium electrode) becomes close to the positive electrode potential assigned to the right jellium. As a consequence of the negative electric potential on the left of atom 1 and the positive one on the right of atom 2, the localization of the voltage drop between atoms 1 and 2 is naturally realized.

In conclusion, a first-principles calculation method using the Lippmann-Schwinger equation was implemented to elucidate electron-transport properties of Na nanowires at finite bias voltages. Employing a nanowire model including the plinths as well as the semi-infinite jellium electrodes, we obtained the nonlinear I-V characteristic and the unusual behavior of the differential conductance. We demonstrated that the negative feature of the differential conductance is explained by the suppression of the transmission peaks due to the electron transportability of the negatively biased plinth. It was also found that the voltage drop preferentially occurs around the negatively biased end of the nanowire, and this finding was discussed in relation to the electronic structure and conduction of the nanowire.

This work was supported by a Grand-in-Aid for COE Reseach (No.~08CE2004) from the Ministry of Education, Culture, Sports, Science and Technology.



\begin{thebibliography}{99}
\bibitem{comment1}The spacial distribution of the LDOS relevant to the peak A (B) shows an eigenmode pattern with a loop (node) at the center of the nanowire. Because of this distinct difference, the peak B is not regarded as a shifted one of the peak A to the higher energy side.
\end{thebibliography}

\end{document}